\documentclass{elsarticle}
\usepackage{graphicx}

\newcommand{\lsim}{\mbox{\raisebox{-1.ex}{$\stackrel{<}{\sim}$}}}
\newcommand{\gsim}{\mbox{\raisebox{-1.ex}{$\stackrel{>}{\sim}$}}}
\newcommand{\vct}[1]{\mbox{\boldmath${#1}$}}

\begin{document}
\begin{frontmatter}

\title{Revised production cross-section of $\gamma$-rays  
in p-p collisions with LHC data
 for the study of
 TeV $\gamma$-ray astronomy}

\author{H. Sato, T. Shibata, R. Yamazaki}


\begin{center}
\address{Department of Physics and Mathemtics, 
  Aoyama-Gakuin University,\\
 Chuo-ku Fuchinobe 5-10-1, 
Kanagawa 252-5258, Japan}
\end{center}

\begin{abstract}
 We present the production cross-section of $\gamma$-rays
 based on data of p-p collisions at the Large Hadron Collider (LHC),
 revising the previous semi-empirical formula mainly for 1) the inelastic
cross-section in  p-p collisions, $\sigma_{\mbox{pp}}(E_0)$,
 and 2) the inclusive $\gamma$-ray spectrum in the forward
 region, $\sigma_{\mbox{pp} \rightarrow \gamma}(E_0, E_\gamma)$.
 We find that the previous cross-section gives a significantly 
 softer spectrum than found in the data of LHC. 
 In this paper, we focus our interest mainly upon the LHC forward (LHCf) 
experiment, giving 
  $\gamma$-ray spectra in 
 the very forward region with the pseudo-rapidity $\eta^*$\,$\gsim$\,8.8
 in the center of mass system (CMS), 
  which have not been reported so far. We also give the pseudo-rapidity
 distribution of charged hadrons with $-3 \le \eta^* \le 3$ obtained by
 ALICE and TOTEM experiments, both with LHC. We find that the revised 
cross-section reproduces quite well 
the accelerator data over the wide energy range from GeV to 30\,PeV
  for projectile protons, corresponding approximately to 100\,MeV to 3\,PeV
 for secondary $\gamma$-rays.
  The production cross-section of $\gamma$-rays
 produced in the forward region is essential for the study of $\gamma$-ray
 astronomy, while not important are those produced in the central region in CMS,  and of much less importance in the backward.
We  discuss also the average  transverse momentum
 of $\gamma$-rays, $\bar{p}_{\mbox{t}}$, and
the average inelasticity transferred to $\gamma$-rays,
  $\bar{k}_\gamma^*$, obtaining that the former increases very slowly 
with $\bar{p}_{\mbox{t}} = 100 \sim 220$\,MeV/c 
for $E_0 = 1\,\mbox{GeV} \sim  26\,\mbox{PeV}$, and the latter
 is almost independent of $E_0$, with $\bar{k}_\gamma^* \approx 1/6,$
while we can not exclude the possibility of a small
 increase of $\bar{k}_\gamma^*$.
\end{abstract}

\end{frontmatter}

\section{Introduction.}
The high energy $\gamma$-ray observations open a new window not only for
  the astronmy, but also for the particle physics, cosmology, which must 
 bring us critical information and hints for the understanding of 
 current questions such as nature of 
the dark matter, mechanism of the $\gamma$-ray burst,  
origin of the highest cosmic-ray (CR), and so forth. In recent years, 
 remarkably developed are techniques in both on-board and ground-based
 telescopes, {\it Fermi}, H.E.S.S., MAGIC, VERITAS, which 
 have reported exciting results for the sky-map of GeV-TeV
 $\gamma$-rays in space, still working continuously, see \cite{R2} for 
 review. 

A new program called CTA (Cherenkov
 Telescope Array) is further progressing in the form of 
 the international collaboration consisting of many scientists from 
Europe, USA and Japan \cite{R28}. 
  It covers very wide energy range of $\gamma$-rays with several tens GeV to
 more than 10 TeV with much higher
 sensitivity than ever achieved, the full operation of which is scheduled
 around 2020. If the CTA program is operated as scheduled, we expect to 
 detect TeV $\gamma$-ray sources more than thousands, 
comparable with those currently observed
 in GeV region, which must bring us surely new aspects for
 the understanding of the universe.

Under these situations, it is quite desireable to have
 a reliable production cross-section of hadron-induced $\gamma$-rays
 with TeV energy or more, while those of electron-induced ones are,
 needless to say, well established on the basis of QED. 
 Stecker \cite{R3} presented the production cross-section of $\pi^0$,
$\sigma_{\mbox{pp} \rightarrow \pi^0}(E_0, E_{\pi^0})$
 with $E_{\pi^0}$\,$\lsim$\,100\,GeV ($E_0$\,$\lsim$\,1\,TeV) 
 in the form of very useful parameterizations in
 1973, and later one of the authors (T.\ S.) revised it 
in the previous paper \cite{R1} (hereafter
  Paper I), applicable for the higher energy region $E_0$\,$\gsim$\,1\,TeV. 
 Nevertheless, the reliability is not always satisfactory, particularly
 in the forward region for  
 $E_\gamma$\,$\gsim$\,1\,TeV ($E_0$\,$\gsim$\,10\,TeV) because of the
 limitation in the accelerator data, having no data on the energy
 spectra in the forward region, while Chacaltaya emulsion chamber
  (EC) group [20, 33] have given them using CR-beams, but poor in
 data quality. 
 
Fortunatelly, the LHC experiment started the operations
 in early 2010, among
 which the LHCf group measured $\gamma$-ray spectra in the very
 forward region with the pseudo-rapidity $\eta^*\ \gsim \ 8.8$ 
 in the CMS \cite{R4}. The principal purpose of
 the LHCf is to find the best simulation code in hadron-interaction
 models, which
  plays a key role in the study of the extensive air shower (EAS) 
 phenomena induced by the ultra-high energy CRs, 
affecting directly the estimation of the primary
 CR energy as well as its  composition. While none of models currently
 available reproduce  satisfactorily the LHCf data according to their
 preliminary studies \cite{R4},
 they will report a revised model soon based on further analyses
  as well as on coming data at $\sqrt{s} = 14$\,TeV, the run of 
which is scheduled in 2014.
 
 Alternatively, the LHCf data also give us crucial information  for the
 study of  hadron-induced $\gamma$-rays in galactic environments, both 
interstellar medium (ISM) and the source of CRs, 
 typically the supernova remnant (SNR). Particularly interesting is the
  production cross-section of $\gamma$-rays with 100\,TeV or more 
in the laboratory system (LS), corresponding to the
 knee energy around PeV. We expect that 
 the knee problem in close connection with  
the acceleration limit of CRs in the SNR may be solved by the CTA program
 through the
 observations of ultra-high energy $\gamma$-rays with $E_\gamma$ $\gsim$
 100\,TeV, possibly much more clearly than those of hadronic components.
 Note that it is still not cleared albeit so many years have passed
 since the discovery of the knee \cite{R30}. This is mainly due to the
 difficulty in observing  the latter components around the
 knee energy by both  direct (balloon and/or satellite) 
 and the indirect (EAS) methods, which have inevitable weaknesses in
 statistics for the former, and in the uncertainty of the composition
 for the latter. 
 
 Now the accelerator data on 
 $\sigma_{\mbox{pp} \rightarrow \gamma}(E_0, E_\gamma)$ being 
 available over extremely wide energy ranges, $E_0\, =\, 1$\,GeV\,$\sim$ 
30\,PeV, it is an easy task to find empirically 
 the cross-section by {\it interpolating} them without asking for 
  complicated models in the meson production.
 While many phenomenological models with QCD
 have been applied for the current simulation codes \cite{R29}, 
one should keep in mind the fact that even the inelastic collision
 cross-section based on QCD, $\sigma_{ \mbox{pp}}(E_0)$,
 is not yet definitely established, 
and much less successful are those for the multiple
 meson production, 
$\sigma_{ \mbox{pp} \rightarrow \pi}(E_0, E_\pi)$.
 In the present paper, we give 
$\sigma_{\mbox{pp} \rightarrow \gamma}(E_0, E_\gamma)$
 with simple parameterization  based on the experimental data with LHC,
 revising slightly the previous one,
 which will be quite useful for the study of 
future $\gamma$-ray astronomy even around 100 TeV or more. 

\section{Cross-sections}
\subsection{Inelastic collision cross-section}

In Paper I, we gave the empirical formula
 for the inelastic cross-section,
 $\sigma _{ \mbox{pp}}(E_0)$,
 in p-p collision which covers the wide energy range from 
the  threshold energy of pion production ($\sqrt{s} \approx 2\,\mbox{GeV}$) to
 the FNAL energy ($\sqrt{s} = 1.8\,\mbox{TeV}$).
 Now we have the LHC data (ATLAS[5], ALICE[23], TOTEM[24])
 on $\sigma _{ \mbox{pp}}$ 
at $\sqrt{s} = 7\,\mbox{TeV}$.
 The LHC energy currently available corresponds to approximately
 26\,PeV proton in the LS, hign enough even for 
100\,TeV $\gamma$-ray astronomy.

 Based on the LHC data, we assume a following empirical 
 form for $\sigma _{ \mbox{pp}}$,

\begin{equation}
 \sigma_{ \mbox{pp}}(E_0) = 
{\it \Sigma}_0(E_0) \Bigl[1 + \sqrt[8]{s/s_0}\,\Bigr],
\end{equation}
\vspace{-3mm}
with
\begin{equation}
{\it \Sigma}_0(E_0) = {\sigma}_0\,(c/v)^{\kappa}
 \Bigl(1 - \mbox{e}^{-E_0/\epsilon_0}\Bigr),
\end{equation}
where $E_0$ and $v$  are the kinetic energy and 
velocity of the
 projectile proton in LS respectively, $\sqrt{s_0} = 156$\,GeV, and 
see the first column ($1 \le A_{ \mbox{T}} \le 4$) of Table 1 
 for $\sigma_0$, $\epsilon_0$,  and $\kappa$. Practically, 
${\it \Sigma}_0 \approx {\sigma}_0$ for $E_0\ \gsim\, 10$\,GeV.

\begin{table*}[!t]
\caption{Numerical values of [$\sigma_0, \epsilon_0, \kappa]$
 appearing in Eqs. (1) and (2) for three cases of target mass,
$1 \le A_{ \mbox{T}} \le 4$, $4 < A_{ \mbox{T}} \le 38$,
 and $38 \le A_{ \mbox{T}}$,
 where $\sigma_0$ is in units of millibarn (mb), and $\epsilon_0$ in GeV.}
\begin{center}
\begin{tabular}{rccc}\\ 
\hline
 &  $1 \le A_{ \mbox{T}} \le 4$  &\  $4 < A_{ \mbox{T}}
 \le 38$  & \ \ \  $38 \le A_{ \mbox{T}}$ \\
\hline
$\sigma_0$\,:  
& 20.0 & 22.3 & 18.5
\\ 
$\epsilon_0$\,:  
& 0.93 & 0.92 & 0.92
\\ 
$\kappa$\,:  
& 2.23 & 2.40 & 2.40
\\ 
\hline 
\\
\end{tabular}
\end{center}
\end{table*}

\begin{figure}[b]
\begin{center}
\end{center}
  \caption{
Inelastic  cross-section $\sigma _{ \mbox{pp}}$ in 
proton-proton collision as a function of the center of mass energy $\sqrt{s}$
([9-11], [5], [23], [24]) 
 with two empirical curves, the present one (solid curve) and
 the previous one (dasshed curve). 
$$
\vspace{5mm}
$$
}
\end{figure}

In Fig.\ 1, we demonstrate 
 $\sigma _{ \mbox{pp}}$ against $\sqrt{s}$ 
 including the LHC data, together with  the 
previous empirical curve
 (dasshed curve),  
  and the present one (solid curve) given by Eq.\ (1)
 with the numerical values summarized in the first column of Table~1. 
  One sees that the previous one
 gives significantly over-estimation in the LHC energy region.

For the inelastic collision between nucleus $i$
 and nucleus $j$, we give the cross-section
 based on the optical model [6, 7, 8],
\begin{equation}
 \sigma_{ij}
 = \pi (a_i^2 + a_j^2)[\ln \chi_{ij} + {E}_1(\chi_{ij})+ 
\gamma_{ \mbox{E}}],
\end{equation}
with
\begin{equation}
 \chi_{ij} = A_i A_j \frac{\sigma_{ \mbox{NN}}}
 {\pi (a_i^2 + a_j^2)}, 
\end{equation}
where $A_i$ $(A_j)$ is the mass number of the projectile (target) nucleus,
$\gamma_{ \mbox{E}}$  the Euler constant (=0.5772),
${E}_1(\chi)$ the exponential integral function, and 
$a_i$ ($a_j$) is related to the nuclear root-mean-square radius
 of the nucleus $i$ ($j$), see \cite{R7} for the explicit value of
 $a_i$ ($a_j$). Here $\sigma_{ \mbox{NN}}$ is given by
 Eq.\ (1), but numerical values of the parameters appearing there depend 
 slightly on the target nucleus $A_{ \mbox{T}}$ as presented
 in Table~1.

\subsection{Production cross-section of $\gamma$-rays}

  Now in the following discussions, we use the natural units with
 $c = 1$ (the speed of light) unless otherwise mentioned
 specifically, and  the asterisk
 attached to variables denotes those in the CMS, or else those in the LS.
 
In Paper I, we assumed that the distributions of 
the energy and transverse momentum in the multiple meson production 
are both given by the exponential function,
 $\exp[-(E_\gamma^*/T_0 + p_{ \mbox{t}}/p_0)]$
 ($E_\gamma^*$: $\gamma$-ray energy,  
$p_{ \mbox{t}}$: its transverse momentum), 
 which are expected from the fire-ball picture \cite{R12}. 
On the other hand in the QCD picture, 
 the energy distribution is given by the 
 algebraical function, $(1-x^*)^m$, expected from the quark dimensional 
 counting approach \cite{R13}, where $x^*$ is the light cone
 momentum fraction, and $m$ relates to the number of quarks
 actively involved in the collision with $m = 3$\,$\sim$\,6 
 practically.

 While  both types of the distribution are 
  equivalent in the central region around $x^* \approx 0$, 
we find that the algebraical-type reproduces rather
 well the LHCf data in the exteremely forward region
 with $x^* \approx 1$ (see Section 3.3). So 
 in the present paper, 
 we assume a following distribution function for the invariant
 production cross-section of $\gamma$-rays, modifying slightly the
 functional form used in Paper I,
\begin{equation}
\frac{1}{\sigma_{ \mbox{pp}}}E_\gamma^* 
\frac{d^3\sigma}{d^3{\vct{p}}_\gamma^*}
= \frac{\bar{N}_\gamma {\it \Theta_{ \mbox{c}}}}{4\pi
 T_{ \mbox{c}}^2}
\frac{(1-x^*)^m}{X^*}
\mbox{e}^{-p_{ \mbox{t}}/p_0},
 \label{eq:form}
\end{equation}
with
\begin{equation}
x^* ={E_\gamma^*}/T_{ \mbox{c}};\ \  
T_{ \mbox{c}} = \sqrt{s}/2 = M_{ \mbox{p}}
 \gamma_{ \mbox{c}},
 \label{eq:XgTc}
\end{equation}
\begin{equation}
X^* = x^* + {\zeta p_{ \mbox{t}}}/{p_0};\ \ 
 p_{ \mbox{t}} = E_\gamma^* \sin \theta^*,
 \label{eq:Xgamma}
\end{equation}
where $\gamma_{ \mbox{c}}$ is the Lorentz factor of
 CMS against the LS, $M_{ \mbox{p}}$ the proton mass,
   and ${\it \Theta}_{ \mbox{c}}$ is the 
  normalization constant.

\begin{table*}[!b]
\caption{Explicit forms of $F_{0}(\tau)$ and $F_{1}(\tau)$,
 and their approximate expressions 
  for $\tau \approx 0$.}
\begin{center}
\begin{tabular}{lc}\\ 
\hline
 explcit forms of $F_{0}(\tau)$ and $F_{1}(\tau)$  &\ \ \ \ \ 
  for  $\tau \approx 0 $ \\
\hline \\ 
{$\displaystyle 
F_0(\tau) = 
\frac{1}{\tau}-\frac{4}{\tau^2}+\frac{12}{\tau^3}
-\frac{24}{\tau^4} + 
\frac{24}{\tau^5} \biggl[1-\mbox{e}^{-\tau}\biggr]$}
 & \ \ \ \ \
{$\displaystyle \approx \ \frac{1}{5}\ -\ \frac{\tau}{30} + 
\frac{\tau^2}{210}$}
\\ \\ 
{$\displaystyle
 F_1(\tau) = 
\frac{1}{\tau^2}-\frac{8}{\tau^3}+\frac{36}{\tau^4}
-\frac{96}{\tau^5} + 
\frac{120}{\tau^6} 
\biggl[1-\Bigl(1+\frac{\tau}{5}\Bigr)\mbox{e}^{-\tau}\biggr]$}
 & \ \ \ \ \
{$ \displaystyle \approx \ 
\frac{1}{30}-\frac{\tau}{105}+\frac{\tau^2}{560}$}
\\ \\ 
\hline 
\\
\end{tabular}
\end{center}
\end{table*}

In Eq.\ (5), we introduce four parameters, 
$[m, \zeta, \bar{N}_{\gamma}, p_0]$, where $m$ corresponds to
 the {\it softness} of the energy-spectrum, $\zeta$ to the
  correlation strength between $E_\gamma^*$ and 
 $p_{ \mbox{t}}$, and 
$\bar{N}_{\gamma}$ is of course the multiplicity of $\gamma$'s, 
 and $p_0$ links to the average transverse
 momentum $\bar{p}_{ \mbox{t}}$ (see Eq.\ [15]).
 In the present work, however, we fix the former two, 
 with $m=4$ and $\zeta=0.02$, expected from the preparatory
 calculations \cite{R31}, and focus our principal work on the 
 determination of the latter two, $[\bar{N}_{\gamma}, p_0]$,
 (practically $[\bar{N}_{\gamma}, \bar{p}_{ \mbox{t}}]$,
 see the beginning of Section 3) by the least square method 
 in fitting with the experimental data.

 In order to make 
 following discussions easy to understand, we introduce a 
 parameter $\tau_{ \mbox{c}}$, 
 corresponding to $\tau_0$ defined by Paper I,
\begin{equation}
 \tau_{ \mbox{c}} = T_{ \mbox{c}}/p_0
 = \sqrt{s}/2p_0,
 \label{eq:tauc}
\end{equation}
\vspace{-7mm}
 and   ${\it \Theta}_{ \mbox{c}}$ is given by
\begin{equation}
\frac{1}{{\it \Theta}_{ \mbox{c}}} =
 \int_0^1 \frac{F_0(\tau_\theta^*)}
{1+\zeta \tau_{\theta}^*} d(\cos \theta^*);\ 
 \tau_\theta^* = \tau_{ \mbox{c}}\sin \theta^*,
 \label{eq:Theta}
\end{equation}
\vspace{-7mm}
with
\vspace{-3mm}
\begin{equation}
F_{\ell}(\tau) = \int_0^1 x^{\ell}
 (1-x)^4 \mbox{e}^{-\tau x} dx,\ \ 
 (\ell = 0, 1).
 \label{eq:Fnu}
\end{equation}
\vspace{-3mm}

In Table 2 we give the explicit forms of $F_{\ell}(\tau)$ $(\ell =0, 1)$, 
 together with those of the approximation around $\tau \approx 0$, and 
 demonstrate the numerical value of 
${\it \Theta}_{ \mbox{c}}$ against 
$\tau_{ \mbox{c}}$ for several choices of $\zeta$
in Fig.\ 2. 
\begin{figure}[!t]
\begin{center}
\end{center}
  \caption{
 Normalization constant ${\it \Theta}_{ \mbox{c}}$ against
 $\tau_{ \mbox{c}}$
 for several choices of $\zeta$.
}
\end{figure}

\subsection{Energy-angular distribution}

Eq.\ (5) is rewritten with $m=4$ as
\begin{equation}
\frac{d^2 N_\gamma}{dE_\gamma^* d{\it \Omega}^*} =
\frac{\bar{N}_\gamma {\it \Theta_{ \mbox{c}}}}
{4\pi T_{ \mbox{c}}}
\frac{(1-x^*)^4}{1+\zeta \tau_{\theta}^*}
  \mbox{e}^{-\tau_{\theta}^* x^*},
\end{equation}
\vspace{-3mm}
 which is useful practically for the comparison with  experimental data
 in the CMS as presented in the next section.

Remembering the invariant phase space, 
$E_\gamma^* dE_\gamma^* d{\it \Omega}^*
 =E_\gamma dE_\gamma d{\it \Omega}$, the above equation gives the
 distribution function in LS as
\begin{equation}
\frac{d^2 N_\gamma}{dE_\gamma d{\it \Omega}} =
\frac{\bar{N}_\gamma
 {\it \Theta_{ \mbox{c}}}
\beta_{ \mbox{c}}^2}
{2\pi M_{ \mbox{p}}}
\frac{(1-x{\it \Gamma}_\theta)^4}
{{\it \Gamma}_\theta +\zeta \tau_\theta}
  \mbox{e}^{-\tau_\theta x}, 
\end{equation}
\vspace{-3mm}
with $x = {E_\gamma}/{E_0}$, and 
\begin{equation}
\tau_\theta = 2 (\gamma_{ \mbox{c}}^2 -1)
(M_{ \mbox{p}}/p_0) \sin \theta,
\end{equation}
\begin{equation}
{\it \Gamma}_\theta = 
2(\gamma_{ \mbox{c}}^2 -1) 
(1- \beta_{ \mbox{c}} \cos \theta),
\end{equation}
note
${\it \Gamma}_\theta \approx 1 + \gamma_{ \mbox{c}}^2 \theta^2$ 
for $E_0 \gg M_{ \mbox{p}}$
 in the forward region, leading to $x \approx x^*$.

\subsection{Average transverse momentum and inelasticity}

In this subsection, we give the average transverse momentum of $\gamma$-rays
 using Eq.\ (11) (or Eq.\ [5]),
 which is a critical parameter in the multiple meson production,
 with a quite stable value, say 150\,MeV/c, almost independent of
 the interaction energy. It is immediately given by 
\begin{equation}
\frac{\bar{p}_{ \mbox{t}}}{p_0} =
{\it \Theta}_{ \mbox{c}}  \int_0^1
\tau_\theta^*  \frac{ F_{1}(\tau_\theta^*)}
{1+\zeta \tau_\theta^*}d(\cos \theta^*).
 \label{eq:ptp0}
\end{equation}

The total energy flow transferred to $\gamma$-rays in the CMS, 
${{\it \Sigma} E}_\gamma^{\,*}$,  is similarly 
 obtained from Eq.\ (11),
\begin{equation}
\frac{{{\it \Sigma} E}_\gamma^{\,*}}
{\bar{N}_\gamma T_{ \mbox{c}}} = 
\frac{\bar{E}_\gamma^*}{T_{ \mbox{c}}} = 
{\it \Theta}_{ \mbox{c}}  \int_0^1 
  \frac{F_{1}(\tau_\theta^*)}
{1+\zeta \tau_\theta^*} d(\cos \theta^*),
\end{equation}
where $\bar{E}_\gamma^*$ is the average energy of 
 $\gamma$-rays in the CMS.
In Fig.\ 3, we present ${\bar{p}_{ \mbox{t}}}/{p_0}$ and
 $\bar{E}_\gamma^*/T_{ \mbox{c}}$ simultaneously 
 against $\tau_{ \mbox{c}}$ for several choices of 
  $\zeta$,  corresponding to Fig.\ 2.

\begin{figure}[!b]
\begin{center}
\end{center}
  \caption{
Numerical values of 
${\bar{p}_{ \mbox{t}}}/{p_0}$ and
$\bar{E}_\gamma^*/T_{ \mbox{c}}$
 against $\tau_{ \mbox{c}}$ for several choices of  
$\zeta$,
 where the former corresponds to the left axis, and
 the latter to the right axis respectively.
$$
\vspace{5mm}
$$
}
\end{figure}

Defining the average inelasticity transferred to $\gamma$-rays in the CMS,
\begin{equation}
{\bar{k}_\gamma^*} = \frac{{{\it \Sigma} E}_\gamma^{\,*}}
{\sqrt{s} - 2M_{ \mbox{p}}},
\end{equation}
we have
\begin{equation}
{\bar{k}_\gamma^*}
 = \frac{\bar{N}_\gamma {\it \Theta}_{ \mbox{c}}}{2} \int_0^1
  \frac{F_{1}(\tau_\theta^*)}
{1+\zeta \tau_\theta^*} d(\cos \theta^*),
\end{equation}
 where we use the approximation $\sqrt{s} \gg 2M_{ \mbox{p}}$
 for the practical purpose.
Assuming ${\bar{k}^*_{ \mbox{tot}}}
= 1/2$ (total inelasticity) and $p_{\pi^0} = 1/3$ (isospin symmetry), we
 expect ${\bar{k}_\gamma^*}=1/6$, which is discussed again in Section 4.

\subsection{Pseudo-rapidity distribution}

Practically in the accelerator data, we often use a variable of
 pseudo-rapidity  $\eta^*$ instead of\ \,$\theta^*,$ defined by
 $\eta^* = - \ln \tan \theta^*/2$. The pseudo-rapidity distribution is
 immediately given by, after integrating with respect to $x^*$ in
 Eq.\ (11), 
\begin{equation}
 \frac{d N_\gamma}{d\eta^*} = 
\frac{\bar{N}_\gamma{\it \Theta}_{ \mbox{c}}}
{2\cosh^2\eta^{*}}
\frac{F_{0}(\tau_\eta^*)}{1+\zeta \tau_\eta^*};\ \ 
\tau_\eta^* = \frac{\tau_{ \mbox{c}}}{\cosh \eta^*}.
\end{equation}

It is also important to see the energy flow,
 $d({\it \Sigma} E_\gamma^*)$, 
  transferred to $\gamma$-rays within ($\eta^*$, $\eta^* + d\eta^*$),  which is given by
\begin{equation}
\frac{d ({\it \Sigma} E_\gamma^*)}{T_{ \mbox{c}} d\eta^*} = 
\frac{\bar{N}_\gamma{\it \Theta}_{ \mbox{c}}}{2\cosh^2\eta^{*}}
\frac{F_{1}(\tau_\eta^*)}{1+\zeta \tau_\eta^*},
\end{equation}
\vspace{-3mm}
see Table 2 for $F_{0}(\tau)$ and $F_{1}(\tau)$.

The energy flow, $d({\it \Sigma} E_\gamma)$, 
 in LS  within ($\eta^*$, $\eta^* + d\eta^*$) 
is easily given by 
\begin{equation}
\frac{d ({\it \Sigma} E_\gamma)}{d\eta^*} = 
\frac{\cosh (\eta^*+\eta_{ \mbox{c}})}
{\cosh \eta^*}
\frac{d ({\it \Sigma} E_\gamma^*)}{d\eta^*},
\end{equation}
\vspace{-3mm}
with $\eta_{ \mbox{c}} = \frac{1}{2}
\ln (1+ \beta_{ \mbox{c}})/(1-\beta_{ \mbox{c}})$, i.e., 
rapidity of the CMS against the LS.
One finds a reasonable relation between the total energy flow
in LS and that in the CMS,  
${{\it \Sigma} E}_\gamma =
 \gamma_{ \mbox{c}} 
{{\it \Sigma} E}_\gamma^{\,*}$, after integrating 
 both sides of Eq.\ (21) over $\eta^*$
  (see Eq.\ [16] for the explicit form of 
${{\it \Sigma} E}_\gamma^{\,*}$) 
because of the forward-backward symmetry in the CMS, 
 leading to $\bar{k}_\gamma \approx \bar{k}_\gamma^*$.

\begin{figure}[!t]
\begin{center}
\end{center}
  \caption{
Three kinds of density distribution (see Eq.\ [22]) 
against the pseudo-rapidity $\eta^*$ 
 in the differential form, a) multiplicity, b)
 energy flow in CMS, and c) energy flow in LS,
 for three energies, $E_0= 10^{11}$, $10^{14}$, and
$10^{17}$eV.
$$
\vspace{3mm}
$$
}
\end{figure}

\begin{figure}
\begin{center}
\end{center}
  \caption{
Three kinds of density distribution against the pseudo-rapidity $\eta^*$
 in the integral form, corresponding to those in the differential
 form presented in Fig.\ 4.
$$
\vspace{3mm}
$$
}
\end{figure}

In Fig.\ 4, we demonstrate three kinds of density distribution 
against $\eta^*$ simultaneously, 
$\rho_{ \mbox{N}}^*$ (multiplicity), 
  $\rho_{ \mbox{E}}^{*}$ (energy flow in CMS), and  
 $\rho_{ \mbox{E}}$ (energy flow in LS), 
  for $E_0 = 10^{11}, 10^{14}, 10^{17}$\,eV, where the vertical
 axis is normalized to unity after integrating over $\eta^*$, 

\begin{equation}
\rho_{ \mbox{N}}^* = \frac{dN_\gamma}{{\bar{N}_\gamma}d\eta^*},\ 
\rho_{ \mbox{E}}^{*} = \frac{d ({\it \Sigma} E_\gamma^{\,*})}
{{{\it \Sigma} E}_\gamma^{\,*} d\eta^*},\ 
\rho_{ \mbox{E}} = \frac{d ({\it \Sigma} E_\gamma)}
{{{\it \Sigma} E}_\gamma d\eta^*}.
\end{equation}

 On the other hand,  we present the integral forms for three
 kinds of density, 
$P_{ \mbox{N}}^*(\ge$$\eta^*)$, 
$P_{ \mbox{E}}^{*}(\ge$$\eta^*)$, and
$P_{ \mbox{E}}(\ge$$\eta^*)$ in Fig.\ 5, 
 corresponding to Fig.\ 4. From these figures, 
  one finds that the energy flow in LS comes from mostly 
those produced in the 
 very forward region in CMS, particularly for higher energy.
 This means that essential is the production energy spectrum of 
$\gamma$-rays in the very forward region in CMS, contrarily 
 not important in the central region, and of much less importance 
  in the backward. 

While the present paper is focussed upon the $\gamma$-ray component
 produced by $\pi^0$-decay, the pseudo-rapidity distributions of
 {\it charged hadrons}, $dN_{ \mbox{ch}}/d\eta^*$, 
 have been extensively studied with accelerator experiments, 
     closely related to $dN_{\gamma}/d\eta^*$ given by Eq.\ (19). 
 So we present the 
 kinematical relation between them rather in detail in Appendix A,
  which is given by, 
assuming $\bar{N}_\gamma \approx \bar{N}_{ \mbox{ch}}$, 
\begin{equation}
\frac{dN_{ \mbox{ch}}}{d\eta^*} \approx
\biggl [ 1 + \frac{1}{2} \biggl(\frac{m_{\pi^0}}{p_0}\biggr)^2
\frac{{\it \Delta}(\eta^*)}{F_0(\tau_{\eta^*})}
\tanh ^2 \eta^*  \biggr]
\frac{dN_{\gamma}}{d\eta^*},
\end{equation}
\vspace{-3mm}
where $m_{\pi^0}$ is the mass of $\pi^0$, and see
 Eq.\ (A12) in the appendix for ${\it \Delta}(\eta^*)$.

\section{Comparison with experimental data}
\label{sec:comparison}

In this section, we determine two parameters appearing in Eq.\ (5),
$[\bar{N}_\gamma, p_0]$, while the other two, $[m, \zeta]$,
 are fixed to
 [4, 0.02] as mentioned in Section 2.2. Now 
 the average transverse momentum $\bar{p}_{ \mbox{t}}$ 
is quite stable, and well established in both accelerator and CR experiments 
with 150\,$\sim$\,200\,MeV/c. So  we use the parameter
 $\bar{p}_{ \mbox{t}}$ in place of $p_0$, which is given by
 Eq.\ (15),
\begin{equation}
 \frac{\bar{p}_{ \mbox{t}}}{p_0} = 
  \int_0^1 \tau_\theta^*
  \frac{F_{1}(\tau_\theta^*)}
{1+\zeta \tau_\theta^*}d(\cos \theta^*)
\Biggl /
\int_0^1  \frac{F_{0}(\tau_\theta^*)}
{1+\zeta \tau_\theta^*} d(\cos \theta^*),
 \label{eq:inel}
\end{equation}
\vspace{-3mm}
while we have to solve numerically the above transcendetal equation with
 respect to $p_0$,
 note that it also appears in $\tau_\theta^* = \tau_{ \mbox{c}}
\sin \theta^* = 
 T_{ \mbox{c}} 
\sin \theta^*/p_0$ in the right-hand side of Eq.\ (24).
 In practice, $p_0$ is easily obtained by the iteration method with the
 initial value of 200\,MeV/c for the set of
 $[E_0, \bar{p}_{ \mbox{t}}]$ 
(or $[T_{ \mbox{c}}, \bar{p}_{ \mbox{t}}]$),
 since $p_0$ is also quite
 stable with 150-250\,MeV/c for $\bar{p}_{ \mbox{t}}$\,=\,
 100-230\,MeV/c (see Figs.\ 3 and 17).

 Explicit values of 
$[\bar{N}_\gamma, \bar{p}_{ \mbox{t}}]$ 
are presented in each figure appearing in the following subsections
 (see also Figs.\ 16 and 17, and Table 3), 
which are obtained by 
 fiting the experimental data with the present empirical curve.

\subsection{The low energy region \rm{(}$E_0 = 1 - 300$\,\rm{GeV})}

\begin{figure}[!t]
\begin{center}
\end{center}
  \caption{
Energy spectrum of $\gamma$-rays at $E_0 = 0.97$\,GeV \cite{R14}.
 Empirical curves are given by Eq.\ (12) after integrating over
 $\cos \theta$.
}
\end{figure}

\begin{figure}
\begin{center}
\end{center}
  \caption{
Energy spectrum of $\gamma$-rays at $E_0 = 23.1$\,GeV 
 \cite{R15} for
 different emission angles. Empirical curves are given by Eq.\ (12).
$$
\vspace{7mm}
$$
}
\end{figure}

From Eq.\ (12), we can obtain easily the energy distribution in 
 LS integrating over $\cos \theta$.
In Fig.\ 6, we compare the empirical one thus obtained
 with the data at $E_0=0.97$\,GeV
 given by Bugg et al.\ \cite{R14}{\footnote{Original data were
 given in the form of the production cross-section for
 $\pi^0$, $\sigma_{ \mbox{pp} \rightarrow \pi^0}
(E_0, E_{\pi^0})$.
 As they gave explicitly the number of events per 25\,MeV 
 energy bin, we converted them into 
$\sigma_{ \mbox{pp} \rightarrow \gamma}
(E_0, E_\gamma)$ by randomly sampling for the
 $\pi^0 \rightarrow 2\gamma$ decay in each energy bin.}}, 
 and find that they are well reproduced with the numerical
 values of 
 $\bar{N}_\gamma$ and $\bar{p}_{ \mbox{t}}$ 
  presented in the figure.

Fidecaro et al.\ \cite{R15} gave the production cross-section of 
$\gamma$-rays for different emission angles at $E_0 = 23.1$\,GeV
 in the LS, which is presented in Fig.\ 7 together with 
 curves expected from Eq.\ (12).
 One finds the agreement is excellent for 
 all emission angles.

In Fig.\ 8, we show the energy distribution with use of the
 Feynman variable, $x_{ \mbox{F}}^*
 = x^* \cos \theta^*$, at two energies, 
$E_0 = 11.5$,  203.7\,GeV [16, 17].
 Empirical curves are obtained 
 by replacing  $x^*$ with $x_{ \mbox{F}}^* \sec \theta^*$
 in Eq.\ (11), where we must take care of the integral
 range for $\cos \theta^*$ with $x_{ \mbox{F}}^* 
\le \cos \theta^* \le 1$. One finds the empirical ones
 reproduce well both data.
In Fig.\ 9, we present the pseudo-rapidity distributions
 at $E_0 = 11.5$, 203.7, and 299.1\,GeV
  [16, 17, 18],
 together with the empirical ones obtained by Eq.\ (19).
 Our numerical curves are again 
in nice coincidence with the data.

\begin{figure}[!t]
\begin{center}
\end{center}
  \caption{
Production cross-section of $\gamma$-rays with use of
 the Feynman scaling variable 
$x_{ \mbox{F}}^*$ at two energies,
 $E_0 = 11.5$, 203.7\,GeV [16, 17].
}
\end{figure}

\begin{figure}
\begin{center}
\end{center}
  \caption{
Pseudo-rapidity distributions at three energies,
$E_0 = 11.5$, 203.7, 299.1\,GeV [16, 17, 18]. Empirical curves are
 obtained by Eq.\ (19).
}
\end{figure}

\begin{figure}[!t]
\begin{center}
\end{center}
  \caption{
Energy spectrum of $\gamma$-rays for several sets of the emission angle
 in the CMS at ISR energies \cite{R19}.
 Empirical curves are given by Eq.\ (11).
$$
\vspace{3mm}
$$
}
\end{figure}

\subsection{The high energy region \rm{(}$E_0 = 0.5 - 200$\,\rm{TeV)}}

\begin{figure}[!b]
\begin{center}
\end{center}
  \caption{
Fractional energy spectrum of $\gamma$-rays obtained by 
Chacaltaya EC experiments \cite{R20} for three energy flow ranges,
 ${{\it \Sigma}E}_\gamma$\,=\,7-10, 10-20, and 20-50\,TeV,
 each corresponding to the average energy for projectile proton,
 $\langle E_0 \rangle =$ 29.2, 45.1, and 101.2\,TeV respectively. 
$$
\vspace{7mm}
$$
}
\end{figure}

In this subsection, we compare the experimental data in TeV
 region with our cross-section given by Eq.\ (5), which are 
 obtained by ISR, FNAL, and the Chacaltaya EC 
 with CR-beams.
In Fig.\ 10, we present the ISR data \cite{R19} with $\sqrt{s} = 30.2$,
 44.7, and 52.7\,GeV, each corresponding to $E_0 = 0.483$, 1.06,
 1.48\,TeV in the LS respectively, where empirical curves are obtained by
 Eq.\ (11). One might worry about some discrepancies appearing 
  in the low energy region, 
$E_\gamma^*$\,$\lsim$\,1\,GeV, but it is not so critical in the
 practice as mentioned in Section 2.5, namely important is 
 only the high energy
 part in the forward region in CMS, see Fig.\ 5.

\begin{figure}[!b]
\begin{center}
\end{center}
  \caption{
Pseudo-rapidity distributions obtained by 
UA7 \cite{R22}, and the
 Chacaltaya EC experiments \cite{R20} . Empirical curves are
 given by Eq.\ (19).
$$
\vspace{5mm}
$$
}
\end{figure}

As presented in Paper I, the Chacaltaya EC data \cite{R20} provide the
 fractional energy spectrum of $\gamma$-rays,
 $f_\gamma = E_\gamma/{{\it \Sigma}E}_\gamma$, in $E_0 = 30-200$\,TeV
 region. The relation between ${{\it \Sigma}E}_\gamma$ and
 $E_0$ is given by assuming the $\gamma$-ray inelasticity,
 $\bar{k}_\gamma$, while we have to take care of the bias-effect
 in EC experiments ($\bar{k}_{\gamma,  \mbox{bias}} = 0.28)$, 
  see Paper I for the detail. In Fig.\ 11, we show the
 $f_\gamma$-spectrum for both data and curves expected from
 Eq.\ (12) after integrating over the emission angle. The agreement is
 quite well within the statistical error.

We present the pseudo-rapidity distribution in Fig.\ 12 obtained 
by  UA7 \cite{R22}, and EC data \cite{R20}, 
where we present the curve expected
 from Eq.\ (19) with $\sqrt{s} = 600$\,GeV. Again we find
 that the present curve reproduces nicely the data.

In Fig.\ 13, we show the pseudo-rapidity distribution of
  charged hadrons obtained by UA5 \cite{R21}, for three
 energies, $\sqrt{s}$\,=\,52.7, 200, and 546\,GeV.
 The numerical curves are given by Eq.\ (23) with the assumption of
 $\bar{N}_{  \mbox{ch}} = \bar{N}_{\gamma}$, 
 taking the $\pi^0 \rightarrow 2\gamma$ decay into account. We find that they are in good agreement
 with the UA5 data.

\begin{figure}[!t]
\begin{center}
\end{center}
  \caption{
Pseudo-rapidity distributions of charged hadrons obtained by UA5 \cite{R21}. 
 Empirical curves are
 given by Eq.\ (23).
$$
\vspace{5mm}
$$
}
\end{figure}

\begin{figure}[!b]
\begin{center}
\end{center}
  \caption{
Energy spectrum of $\gamma$-rays obtained by LHCf at
 $\sqrt{s} = 7$\,TeV for two sets of
 [${\it \Delta}\eta^*$, ${\it \Delta}\phi^*$],
 (a) [$\eta^*$\,$>$\,10.94, 360$^\circ$] and
 (b) [8.81\,$<$\,$\eta^*$\,$<$\,8.99, 20$^\circ$], 
 where separately presented are two detectors, Arm1 (open circle)
 and Arm2 (filled circle), each with the scintillation fiber and
 the silicon strip respectively.
$$
\vspace{3mm}
$$
}
\end{figure}

\subsection{The LHC energy region \rm{(}$E_0 = 0.4 - 30$\,\rm{PeV)}}

Now we compare our production cross-section with the LHC data
 most recently reported, while the final goal with $\sqrt{s} = 14$\,TeV
 will be available around 2014.
 LHCf group [4] present recently the energy spectra of $\gamma$-rays
 in the very forward region,  $\eta^*\, \gsim\ 8.8$, 
 at $\sqrt{s} = 7$\,TeV,
 corresponding to $E_0 = 26$\,PeV in the LS. 
Let us apply our formula given by Eq.\ (11) for LHCf data, and
 estimate  $[\bar{N}_\gamma, \bar{p}_{ \mbox{t}}]$.


In Fig.\ 14, we give the energy spectra at $\sqrt{s}=7$\,TeV,
 for two sets of [${\it \Delta}\eta^*$, ${\it \Delta}\phi^*$],
 (a) [$\eta^*$\,$>$\,10.94, 360$^\circ$]  and (b)
  [8.81\,$<$\,$\eta^*$\,$<$\,8.99, 20$^\circ$], where
 two curves from our empirical cross-section are presented together.
 We find that they reproduce well the experimental data in spectral shape, 
 but the absolute value of $\bar{N}_\gamma$ (average photon yield) 
   is of approximately 20\% difference between them,
 53.4 for (a) and 68.0 for (b), while the latter is consistent 
 with 71.5 expected from TOTEM with 
the pseudo-rapidity distribution of charged hadrons (see 
 Fig.\ 15), assuming $\bar{N}_\gamma
 \approx \bar{N}_{ \mbox{ch}}$.

\begin{figure}[!t]
\begin{center}
\end{center}
  \caption{
Pseudo-rapidity distribution of charged hadrons obtained by UA5 \cite{R21},
 ALICE \cite{R23},
 and TOTEM \cite{R24} at $\sqrt{s} = 900$\,GeV, 2.36\,TeV, and 7\,TeV,
 where curves are given by Eq.\ (23), assuming $\bar{N}_\gamma
 \approx \bar{N}_{ \mbox{ch}}$.
$$
\vspace{3mm}
$$
}
\end{figure}

In addition to the energy spactra of $\gamma$-rays in the forward region
 obtained by the LHCf group, 
UA5 \cite{R21}, 
ALICE \cite{R23} and TOTEM [24] present the pseudo-rapidity
 distribution of
  charged hadrons in the central region as shown in Fig.\ 15, 
 covering the energies $\sqrt{s}$\,=\,900\,GeV, 2.36\,TeV, and 7\,TeV,
where numerical curves are obtained by Eq.\ (23) with the assumption of 
  $\bar{N}_\gamma \approx \bar{N}_{ \mbox{ch}}$.
  One finds that they are well 
 in consistent with the experimental data, particularly interesting is that the
 concave shape around $\eta^* \approx 0$ is nicely reproduced.

\begin{table*}[!b]
\caption{Summary of [$\bar{N}_\gamma,\,\bar{p}_{ \mbox{t}}$], 
where LHCf-1 corresponds to the
 data with [$\eta^*$\,$>$\,10.94,\ ${\it \Delta}\phi^*$\,=\,$360^{\circ}$], 
 and LHCf-2 to those with
[$8.81$\,$<$\,$\eta^*$\,$<$\,8.99,\ ${\it \Delta}\phi^*$\,=\,$20^{\circ}$], and
 $\lq \lq (n)$" $(n=0, 1, 2, \ldots)$ appearing in the column of
 $E_0$ denotes $\lq \lq 10^n$". Fig.\,13 and Fig.\,15 with asterisk mark 
  correspond to the data from the pseudo-rapidity distribution of charged
 hadrons, assuming $\bar{N}_\gamma \approx \bar{N}_{ \mbox{ch}}$
 and   $\bar{p}_{ \mbox{t}}$($\gamma$) $\approx$ 
  $\bar{p}_{ \mbox{t}}$($\pi$)/2.
}
\begin{center}
\begin{tabular}{rrcccc}\\ 
\hline
Figure : &  Data reference  & $\sqrt{s}$\,(GeV) & $E_0$\,(GeV) 
& $\bar{N}_\gamma$ &
$\bar{p}_{ \mbox{t}}$\,(MeV/c) \\
\hline
Fig.\,6\ :& Bugg et al.\,[14] & 2.31 & 0.97(0) & 0.32 $\pm$ 0.11 & 102 $\pm$ 32
\\ 
Fig.\,7\ :& Fidecaro et al.\,[15] & 6.85 & 2.31(1) & 3.50 $\pm$ 0.28 & 137 $\pm$\ \ 9 
\\ 
Fig.\,8\ :& Jager et al.\,[16] & 5.02 &1.15(1) & 1.68 $\pm$ 0.27 & 120 $\pm$ 16
\\ 
        & Jager et al.\,[17] & 19.7 &2.04(2) & 8.25 $\pm$ 0.93
& 167 $\pm$ 16  
\\ 
Fig.\,9\ :&            [16] & 5.02 &1.15(1) & 2.33 $\pm$ 0.20 
& 125 $\pm$\ \ 9
\\
        &               [17] & 19.7 & 2.04(2) & 7.45 $\pm$ 0.74 
& 167 $\pm$ 18
\\ 
       & Shenger et al.\,[18] & 23.7 & 2.99(2) & 7.75 $\pm$ 0.70
& 155 $\pm$ 16
\\ 
Fig.\,10 :& ISR\,[19]& 30.2& 4.83(2) &  7.80 $\pm$ 0.81
& 141 $\pm$ 15  
\\
        &           & 44.7 &1.06(3)  &  9.02 $\pm$ 1.09
& 136 $\pm$ 17 
\\ 
        &           & 52.7 &1.48(3)&  9.65 $\pm$ 1.03 
& 138 $\pm$ 15
\\ 
Fig.\,11\ :& Chacaltaya\,[20]
                    & 234.& 2.92(4) & 23.0 $\pm$ 4.17
& 189 $\pm$ 34
\\
  &        & 291. & 4.51(4) & 24.8 $\pm$ 4.92
& 190 $\pm$ 38 
\\ 
        &      & 436. & 1.01(5)  & 35.0 $\pm$ 4.49 
& 229 $\pm$ 29
\\ 
Fig.\,12\ :& [20], UA7\,[22]& 615.& 2.04(5) & 30.4 $\pm$ 3.52 
& 211 $\pm$ 17
\\ 
Fig.\,13$^*$:& UA5\,[21]& 52.7& 1.48(3)& 12.2 $\pm$ 2.55 
& 146 $\pm$ 31
\\
        &          & 200. & 2.13(4) & 19.2 $\pm$ 3.05 
& 170 $\pm$ 36
\\ 
        &          & 546. &1.59(5)  & 25.3 $\pm$ 3.28 
& 183 $\pm$ 30
\\ 
Fig.\,14\ :& LHCf-1\,[4]&7000&2.61(7) & 53.4 $\pm$ 5.58 
& 234 $\pm$ 25
\\
        & LHCf-2\,[4]&7000&2.61(7)  & 68.0 $\pm$ 7.02
& 234 $\pm$ 25 
\\
Fig.\,15$^*$:& [24], [23], [21]&900.&4.31(5) & 31.9 $\pm$ 7.01
& 183 $\pm$ 30 
\\
   & [24], ALICE\,[23] &2360&2.96(6)& 46.9 $\pm$ 9.87 
&203 $\pm$ 20
\\ 
   & TOTEM\,[24] &7000 &2.61(7) & 71.5 $\pm$ 15.0 
& 222 $\pm$ 20
\\ 
\hline 
\\
\end{tabular}
\end{center}
\end{table*}

\subsection{The multiplicity and the average transverse momentum}

In Figs.\ 6-15, we present explicitly the 
 numerical sets of $[\bar{N}_\gamma, \bar{p}_{ \mbox{t}}]$
 in the extremely wide energy ranges, $E_0$ = 1\,GeV $\sim$ 26\,PeV,
 which are summarized all together in Table 3.

 For the $\gamma$-ray astronomy, practically the most essential is 
the {\it total} production cross-section
 of $\gamma$-rays, $\bar{N}_\gamma \times \sigma_{ \mbox{pp}}$, 
  no matter how the emission-angle $\theta^*$ 
(or the transverse momentum ${p}_{ \mbox{t}}$) 
 appears in the functional form of the cross-section.
 After Stecker \cite{R3} summarized it in 1973,  we revised it in Paper I
 with the data covering TeV region but without LHC data.

\begin{figure}[!t]
\begin{center}
\end{center}
  \caption{
The total production cross-section of $\gamma$-rays,
 $\bar{N}_\gamma \sigma_{ \mbox{pp}}$, 
 against the proton kinetic energy $E_0$,
 where those compiled by Stecker \cite{R3} (filled squares) are 
 plotted together. The present curve (solid curve) is obtained by
 Eq.\ (1) for  $\sigma_{ \mbox{pp}}$ and Eq.\ (25) for 
  $\bar{N}_\gamma$ respectively,
 where the previous one (dashed curve) is also presented.
$$
\vspace{5mm}
$$
}
\end{figure}

Let us present
$\bar{N}_\gamma\sigma_{ \mbox{pp}}$ against $E_0$
 in Fig.\ 16 with LHC data, using Eq.\ (1) for 
$\sigma_{ \mbox{pp}} (E_0)$, where we 
 give a solid curve obtained by the following empirical form for
 $\bar{N}_\gamma$, 
\\
$$
\vspace{2mm}
\bar{N}_\gamma (E_0) = \bar{N}_0 \hat{E_0}^{0.115}
\biggl[1 - \mbox{exp}\Bigl(-0.47 \sqrt{\hat{E_0}}\,\Bigr)\biggr],
 \eqno{\rm (25a)}
$$
$$
\vspace{5mm}
 \bar{N}_0(E_0) = 8.80 \times 
\biggl[1 - \mbox{exp}\Bigl(-0.15 \sqrt[4]{\hat{E_0}}\,\Bigr)\biggr],
 \eqno{\rm (25b)}
$$
with $\hat{E_0} = E_0 - 2m_\pi$ in GeV, 
and presented together is a dashed curve from the previous
 parametrization \cite{R1} for $\bar{N}_\gamma (E_0)$.
  We plot also $\bar{N}_\gamma\sigma_{ \mbox{pp}}$ (open circles)
 expected from 
 the pseudo-rapidity distribution of charged hadrons, assuming 
$\bar{N}_\gamma \approx \bar{N}_{ \mbox{ch}}$
 (see Figs.\ 13 and 15).

One finds that the previous one 
  gives significantly over-estimation in PeV region, and the present
 one reproduces nicely the experimental points in the
 very wide energy range, $E_0$ = 1\,GeV $\sim$ 26\,PeV.

\begin{figure}[!t]
\begin{center}
\end{center}
  \caption{
Average transverse momentum of $\gamma$-rays against $E_0$,
where we present also those expected from charged pions (open circles) 
[37]   with the assumption of
  $\bar{p}_{ \mbox{t}}$($\gamma$) $\approx$ 
  $\bar{p}_{ \mbox{t}}$($\pi$)/2.
$$
\vspace{5mm}
$$
}
\end{figure}

In Fig.\ 17, we give the average transverse momentum, 
$\bar{p}_{ \mbox{t}}$, against $E_0$, together with
 the empirical curve given by 
$$
\bar{p}_{ \mbox{t}} (E_0) = 
\bar{p}_0 \hat{E_0}^{0.0286}
\biggl[1 - \mbox{exp}\Bigl(-1.156\sqrt[4]{\hat{E_0}}\,\Bigr)\biggr],
\eqno{(26)}
$$
with
$$
\vspace{3mm}
\bar{p}_0 = m_\pi c = 140\,\mbox{MeV/c}, 
$$
while it is of little interest for the $\gamma$-ray astronomy,
 but important
 for the study of shower phenomena in the atmosphere.
 In Fig.\ 17 we plot also the half of the transverse momentum of
 the charged pions [37], assuming 
 $\bar{p}_{ \mbox{t}}$($\gamma$) $\approx$ 
  $\bar{p}_{ \mbox{t}}$($\pi$)/2.
 One finds that $\bar{p}_{ \mbox{t}}$ increases
 slowly with $E_0$, as given by Eq.\ (26).

\section{Discussions}

In the present paper, 
interpolating  experimental data nowadays covering the very wide energy range 
from GeV to 30\,PeV, we have focussed 
 our work on  the construction of the semi-empirical formula
 for the inclusive production cross-section of $\gamma$-rays, 
$\sigma_{ \mbox{pp} \rightarrow \gamma}(E_0, E_\gamma)$, 
without asking for the cumbersome QCD-based models, and 
 find that it reproduces excellently the machine data over the
 very wide energy ranges.

 The present simple parameterization in the formula 
should be compared to the 
 simulation codes currently available in the CR community, which are usually 
very complicated, patching different models separately in low and high energy
 regions, and {\it heavy} in the sense that they are constructed so that 
 all the components ($\pi^\pm, \pi^0, K^\pm, \ldots $) in both soft
 (small $q^2$) and hard (large $q^2$) processes simultaneously match with
 the accelerator data.
 So it is not an easy task for physicists other than a
 developer of simulation code to improve it freely, by contrast with
 the  empirical formula, quite easy to touch the parameters appearing there.
Of course one should keep in mind that our approach (interpolation
 method) is not valid
 for the study of extremely high energy shower phenomena 
 in the atmosphere, say $\gsim$ 10$^{18}$\,eV, where even the LHC can not
 cover, resulting in the need of some theoretical models in order to
 extrapolate the LHC data much higher, while the present parameterization
 is valid enough for
 the future $\gamma$-ray astronomy, even up to PeV-$\gamma$ observation.

\begin{figure}[!t]
\begin{center}
\end{center}
  \caption{
The average inelasticity, $\bar{k}_\gamma^*$ transferred to 
$\gamma$-rays against $E_0$,
 where also 
plotted are those expected from the charged hadrons (filled squares) 
with the assumption of 
$\bar{k}_{\gamma}^* \approx \bar{k}_{ \mbox{ch}}^*/2$.
$$
\vspace{5mm}
$$
}
\end{figure}

We have concentrated our interest upon two parameters, 
 $\bar{N}_\gamma$ and $\bar{p}_{ \mbox{t}}$, 
particularly on the former.
  As mentioned often, the multiplicity $\bar{N}_\gamma$ 
 plays an essential role for the study of $\gamma$-ray astronomy,
 appearing always  in the form of the total production cross-section,
 $\bar{N}_\gamma(E_0) \hspace{-0.7mm} \times \hspace{-0.7mm}
 \sigma_{ \mbox{pp}}(E_0)$ as presented in Fig.\ 16.

 Alternatively, $\bar{N}_\gamma$ is important also for the study of
 the shower phenomena in the atmosphere, appearing in 
 the (total) inelasticity $\bar{k}_{ \mbox{tot}}$
 ($\approx \bar{k}_{ \mbox{tot}}^*$). 
So let us present $\bar{k}_\gamma^*$ (see Eq.\ [18]) transferred to
 $\gamma$-rays against $E_0$ in Fig.\ 18, where also plotted
 are those (filled squares) expected from charged hadrons
 (see Figs.\ 13 and 15), assuming
${\bar{k}_\gamma^*}$ $\approx$  $\bar{k}_{ \mbox{ch}}^*$/2.
An error-bar attached to each square comes from 
 statistical ones to  $\bar{N}_\gamma$ 
as  presented in Fig.\ 16. One finds approximately
  ${\bar{k}_\gamma^*}$ $\approx$ 1/6 as a whole, 
 almost independent of $E_0$, while it is as small as
0.1-0.17 in the low energy region $E_0$ $\lsim$ 1\,TeV, and as large as
 0.15-0.2
 in the high energy region $\gsim$ 10\,TeV, indicating a small increase as
 the energy gets higher. But we reserve the conclusion for future
 studies, either constant or the increase.
\\ 

It has been well-known that the attenuation length
 ${\it \Lambda}$ for the intensity of CR hadronic components
 in the atmosphere is given by ${\lambda}/{\it \Lambda} =
 {1 - \langle(1-k_{ \mbox{tot}})^\beta\rangle}$ \cite{R25}, where 
 $\lambda$ is the collision length,
 and  $\beta$  the index of the integral primary CR spectrum with 
  $\sim$\,1.8. Experimentally we have
${\it \Lambda}/{\lambda} \approx 1.5$,  with for instance 
 $\lambda \approx 70$\,g/cm$^2$ and ${\it \Lambda} \approx 100$\,g/cm$^2$
  \cite{R27}, leading to
 $\bar{k}_{ \mbox{tot}}$ $\approx$ 1/2
 in the energy region $E_0$ $\lsim$ 100\,TeV, assuming the uniform distribution
 in ${k}_{ \mbox{tot}}$, while not yet clear in the
 air shower region $E_0$ $\gsim$ 10\,PeV.
 Anyway the present result is not inconsistent with the common understanding
 in the inelasticity expected from the attenuation of CRs in the
 atmosphere.

Finally we address further two open problems in the present paper; 
(1) the nucleus effect of proton-nucleus (p-A) and/or nucleus-nucleus (A-A) 
collisions for the production cross-section of $\gamma$-rays in
 p-p collisions, and (2) 
the applicability of the present
 empirical cross-section for the galactic phenomena other than
   emissions of the  hadron-induced $\gamma$-rays, particularly for those of
 the electron-positrons coming from $\pi$$-$$\mu$$-$$\mbox{e}$ decays. 

First, for the problem (1), one should remember that the 
effective $\gamma$-rays produced by the nuclear interaction in the galactic
 environments (either in ISM or in SNR) are only those produced in the forward
 region in the CMS, while not important are those in the central and backward
 regions. The effect of the plural interactions inside the nucleus appears only in the latter regions. In fact, it has been experimentally well-known that the
 difference between those produced by p-p and those by p-A (A-A) collisions
 appears
 only in the latter regions, while they are well in coincidence with each other in the forward region. This is the reason why we present Fig. 5, stressing 
  in the present paper how essential are 
 the $\gamma$-rays produced in the forward region, 
in contrast  not important in the central region, and of much less
 importance in the backward.

 Practically, of course, we need the production cross-section of $\gamma$-rays
 for p-A (A-A) collisions, as there exist additionally 
helium and heavier components in CRs (projectiles) 
as well as the helium gas in the ISM (targets), while  
 unfortunately we have not yet a reliable model nowadays for the 
p-A (A-A) collisions.
 We have used the modified wounded-nucleon model of
 Gaisser and Schafer [34] in our past calculations [26, 27], where we  
  mention that the uncertainty in the nucleus effect 
is of the second order for the practice. In order to see 
the uncertainty,  we have introduced so called the
 $\lq \lq$enhancement factor" defined by 
$\epsilon_{ q} =
 q_{ \mbox{all}}/q_{ \mbox{pp}}$, 
taking the energy dependence into account, where 
  $q_{ \mbox{all}}$ is the emissivity of $\gamma$-rays in 
the galaxy produced by all kinds of nuclear interactions with p-p, p-A and A-A,
 and $q_{ \mbox{pp}}$ by those with  p-p only.
 For instance, $\epsilon_{ q}$\,=\,1.54\,(Gaisser-Shafer [34]),
 1.50\,(Cavallo-Gould [35]), 1.60\,(Stephens-Badhwar [36]),
 and 1.53\,(Shibata et al.\ [26]), indicating that the difference in 
 the choice of nucleus interaction model is not so significant
 as compared to that in the choice of the propagation model. 
 These results tell us also that the
 procedure in the calculation of the CR propagation becomes 
 quite simple by the use of the enhancement factor $\epsilon_{ q}$.

Second, for the problem (2), indeed we do not touch
 the intermediate meson, $\pi^0$, except the pseudo-rapidity distribution
 of charged hadrons (Figs.\ 13 and 15), but the decay products $\gamma$'s 
 only, having no interest in the intermediate mesons, either via
 $\pi^0$ or via heavier $\eta, \eta^\prime$, etc.. 
However, as long as
 focussing on the decay products such as electrons and/or neutrinos from
 muons, the present empirical form for $\gamma$'s,  
 $\sigma_{ \mbox{pp} \rightarrow \gamma}(E_0, E_\gamma)$, 
 is valid also for muons produced via pions, 
$\sigma_{ \mbox{pp} \rightarrow \mu}(E_0, E_\mu)$,
 while we have to take care of the mass difference between
 photon and the muon. This is because both decays, 
$\pi^0 \rightarrow \gamma +
 \gamma$ and $\pi^{\pm} \rightarrow \mu^{\pm} + \nu\,(\bar{\nu})$,  are
 isotropic two-body decays in the pion rest system, 
leading to the same kinematics in $\gamma$ and
$\mu^{\pm}$ but different mass. Namely, we do not need the information
 of the intermediate pions also in the case of the muon production
 cross-section. This fact tells us that we have 
  model-independently a kinematical relation between the emissivity
 of $\gamma$-rays and that of (secondary) electron-positrons in the
 galaxy, 
 detail of which will be reported elsewhere
in connection with the galactic electron-positron spectrum.

In the near future,  we will apply the present cross-section, 
$\sigma_{ \mbox{pp} \rightarrow \gamma}(E_0, E_\gamma)$,
 for the observational data in TeV region currently available on
 both diffused $\gamma$-rays and those from the source, while
 one of the authors (T.\ S.) have studied the former components in 
{\it Fermi}
 energy region, 100\,MeV $\sim$ 100\,GeV \cite{R26}, \cite{R27},
 using the old parameterization in
 $\sigma_{ \mbox{pp} \rightarrow \gamma}(E_0, E_\gamma)$.

\section{Acknowledgements}

We thank T. Sako (Nogoya University) for giving us valuable
 information and comments on LHCf experiments. We are also
 grateful to CTA-Japan member for valuable discussions in
 the internal meetings.

\begin{center}
Appendix A: \ \  
 Kinematical relation between $\gamma$ and $\pi^0$ angular-distributions
\end{center}

The kinematical relation  between $\gamma$ and $\pi^0$ energy-distributions
 was studied by Sternheimer \cite{R32}, where 
  he assumed that the opening angle, $\alpha^*$, of two $\gamma$'s
  from the decay of high energy $\pi^0$ 
 is so small that the emission angle of $\gamma$ is approximately equal
 to that of $\pi^0$, $\theta_\gamma^* \approx \theta_{\pi^0}^*$. But this
 approximation seems to be too rough to transform practically from 
 the pseudo-rapidity distribution of $\gamma$'s to that of $\pi^0$'s,
 namely ${d N_\gamma}/{d\eta^*} = 2 {d N_{\pi^0}}/{d\eta^*}$
 in his approximation.
 In this appendix we present a more realistic relation between them.
 
 Let us consider a $\pi^0$ produced by p-p collision with
 $(E_{\pi^0}^*, \theta_{\pi^0}^*, \phi_{\pi^0}^*)$,
 each denoting the energy, emission angle, and the azimuthal angle
 in the CMS respectively, and disintegrate into two $\gamma$'s with 
$(E_{1^*}^*, \theta_{1}^*, \phi_{1}^*)$ and 
$(E_{2^*}^*, \theta_{2}^*, \phi_{2}^*)$, where individual angles
 are those against the collision axis. We define further
 two sets of angles, $({\it \Theta}_1^*, {\it \Phi}_1^*)$ and
$({\it \Theta}_2^*, {\it \Phi}_2^*)$, which are 
the emission angles and the zenith angles
 of two $\gamma$'s against the {\it moving direction} of $\pi^0$.
 In this appendix we put $c=1$ for the sake of simplicity.

 We have following relations in these variables
$$
E_1^* + E_2^* = E_{\pi^0}^*,
\eqno{\rm (A1a)}
$$
$$
\vspace{4mm}
{\it \Theta}_1^* + {\it \Theta}_2^* = \alpha^*,
\eqno{\rm (A1b)}
$$
$$
E_1^*\sin {\it \Theta}_1^* = E_2^*\sin {\it \Theta}_2^*,
\eqno{\rm (A1c)}
$$
and
$$
\cos \theta_1^* = \cos {\it \Theta}_1^* \cos \theta_{\pi^0}^*
     + \sin {\it \Theta}_1^* \sin \theta_{\pi^0}^* \cos {\it \Phi}_1^*.
\eqno{\rm (A2)}
$$

Integrating over ${\it \Phi}_1^*$ for both sides of Eq.\ (A2), we have 
$$
\langle \cos \theta_1^* \rangle 
 = \cos {\it \Theta}_1^* \cos \theta_{\pi^0}^*,
\eqno{\rm (A3)}
$$
and hereafter we omit angle brackets, $\langle \cdots \rangle$, 
in the left-hand side for the simplicity, as we are not interested
 in the azimuthal component.

Remembering a well-known relation in $\pi^0$-decay \cite{R33}
$$
 \alpha^* \approx \frac{m_{\pi^0}}{\sqrt{E_1^* E_2^*}\ }
 = \frac{m_{\pi^0}}{\sqrt{E_1^* (E_{\pi^0}^* -E_1^*)}},
\eqno{\rm (A4)}
$$
for $E_{\pi^0}^* \gg m_{\pi^0}$, we obtain, from Eqs.\ (A1a)-(A1c),
$$
{\it \Theta}_1^* \approx \frac{m_{\pi^0}}{E_{\pi^0}^*}
        \sqrt{\frac{E_{\pi^0}^*}{E_1^*}-1} ,
$$
namely
$$
\vspace{5mm}
\cos {\it \Theta}_1^* \approx 1 -
{\omega}(E_{\pi^0}^*, E_1^*),
$$
with
$$
\vspace{5mm}
{\omega}(E_{\pi^0}^*, E_1^*)
 = 
\frac{1}{2} \biggl( \frac{m_{\pi^0}}{E_{\pi^0}^*} \biggr)^2
     \biggl( \frac{E_{\pi^0}^*}{E_1^*}-1 \biggr). 
$$
Now we have from Eq.\ (A3),
putting $\theta_\gamma^* \equiv \theta_1^*$ and 
$E_\gamma^* \equiv E_1^*$,
$$
\cos \theta_{\pi^0}^*
\approx \Bigl \{1 + {\omega}(E_{\pi^0}^*, E_\gamma^*)
\Bigr \}\cos \theta_{\gamma}^*.
\eqno{\rm (A5)}
$$

Here we have to take care of the above expansion with respect to $\omega$,
 which is based on the approximation with $\theta_{\pi^0}^* \gg \alpha^*$, 
while Sternheimer assumed $\alpha^* \approx 0$  \cite{R32}. This means
 the energy range of $\gamma$ 
is limitted within
($E_{-}^*, E_{+}^*$) for the fixed energy of $\pi^0$, $E_{\pi^0}^*$,
 which are obtained  from   
  Eq.\ (A4) with $\theta_{\pi^0}^* \ge \alpha^*$, and given by 
$$
  E_{\pm}^* = [1 \pm B(E_{\pi^0}^*, \theta_{\pi^0}^*)]E_{\pi^0}^*/2,
\eqno{\rm (A6)}
$$
with
$$B(E_{\pi^0}^*, \theta_{\pi^0}^*) = 
\sqrt{1 - (2m_{\pi^0}/E_{\pi^0}^*\theta_{\pi^0}^*)^2}.
\eqno{\rm (A7)}
$$

Now, we define the energy-angular distribution function of $\pi^0$
 with $t_{\pi^0}^* \equiv \cos \theta_{\pi^0}^* $
$$
\vspace{5mm}
n_{\pi^0}(E_{\pi^0}^*, t_{\pi^0}^*) \equiv
 \frac{d^2 N_{\pi^0}}{dE_{\pi^0}^* dt_{\pi^0}^*},
$$
 and thus the energy-angular distribution function of $\gamma$ is 
given by, putting $t_{\gamma}^* \equiv \cos \theta_{\gamma}^*$, 
$$
n_{\gamma}(E_{\gamma}^*, t_{\gamma}^*) = 2
  \int_{E_\gamma^*}^{T_{ \mbox{c}}}
\frac{dE_{\pi^0}^*}{p_{\pi^0}^*}\int_{-1}^{1} dt_{\pi^0}^*
\delta \Bigl[ t_{\pi^0}^* - \Bigl \{ 1  +
{\omega}(E_{\pi^0}^*, E_\gamma^*)\Bigr \} t_\gamma^*\Bigr]
n_{\pi^0}(E_{\pi^0}^*, t_{\pi^0}^*),
\eqno{\rm (A8)}
$$
 and $T_{ \mbox{c}}$ is the maximum energy of ${\pi^0}$
 given by Eq.\ (6) in the text.

For $E_{\pi^0}^* \gg m_{\pi^0}$ (equivalently ${\omega} \ll 1$),
 we have 
 with use of Eq.\,(A5) 
$$
n_{\pi^0}(E_{\pi^0}^*, t_{\pi^0}^*) \approx
 n_{\pi^0}(E_{\pi^0}^*, t_{\gamma}^*) + 
{\omega}(E_{\pi^0}^*, E_\gamma^*) t_\gamma^*
     \frac{\partial}{\partial t_\gamma^*}
     n_{\pi^0}(E_{\pi^0}^*, t_{\gamma}^*),
$$ 
 and  integrating over $E_{\gamma}^*$ for both sides of Eq.\ (A8) 
in order to obtain the angular distribution, we obtain
$$
n_{\gamma}(t_{\gamma}^*) \approx 2 n_{\pi^0}(t_{\gamma}^*) + 
2 t_{\gamma}^* \int_0^{T_{ \mbox{c}}}dE_{\gamma}^*
\int_{E_{\gamma}^*}^{T_{ \mbox{c}}}
\frac{dE_{\pi^0}^*}{p_{\pi^0}^*}
{\omega}(E_{\pi^0}^*, E_\gamma^*)
     \frac{\partial}{\partial t_\gamma^*}
     n_{\pi^0}(E_{\pi^0}^*, t_{\gamma}^*),
\eqno{\rm (A9)}
$$ 
where one has to take care of the kinematical constraints in
 $(E_{\pi^0}^*, E_{\gamma}^*)$ given by Eqs.\ (A6) and (A7)
 for the practical integrations.

Now we use the approximation given by Sternheimer for
 $n_{\pi^0}$ appearing in the integrand in Eq.\ (A9), 
$$
\vspace{2mm}
n_{\pi^0}(E_{\pi^0}^*, t_{\gamma}^*) \approx
-\frac{1}{2}p_{\pi^0}^* \frac{\partial}{\partial E_{\pi^0}^*}
n_{\gamma}(E_{\pi^0}^*, t_{\gamma}^*),
$$
note that the second iteration for $n_{\pi^0}$ is negligible
 as shown in Fig.\ 19, and next exchange the order of integrations for
 $E_\gamma^*$ and $E_{\pi^0}^*$, taking care of the kinematical 
constraints mentioned before
$$
 \int_0^{T_{ \mbox{c}}}dE_{\gamma}^*
\int_{E_{\gamma}^*}^{T_{ \mbox{c}}}
dE_{\pi^0}^* \  \Longrightarrow \ 
\int_{E_{\eta^*}}^{T_{ \mbox{c}}}dE_{\pi^0}^*
 \int_{E_-^*}^{E_+^*}dE_{\gamma}^*
$$
with
$$
\vspace{3mm}
E_{\eta^*} = 2m_{\pi^0}/\theta_{\gamma}^* =
 m_{\pi^0}/\tan^{-1}(\mbox{e}^{-\eta^*}).
$$
The second term in the right-hand side of Eq.\ (A9) is thus
 given by
$$
-\frac{m_{\pi^0}^2 }{2T_{ \mbox{c}}}
\int_{x_{\eta^*}}^1
\frac{dx}{x}\biggl(\ln \frac{1+B_{x,\eta^*}}{1-B_{x,\eta^*}}
 - B_{x,\eta^*}\biggr)
\frac{t_{\gamma}^*\partial^2}{\partial x \partial t_{\gamma}^*}
n_{\gamma}(T_{ \mbox{c}}x, t_{\gamma}^*),
\eqno{\rm (A10)}
$$
with
$$
B_{x,{\eta^*}} = \sqrt{1-\biggl(\frac{\ x_{\eta^*}}{x}\biggr)^2};\ \ 
x_{\eta^*} = \frac{m_{\pi^0}/T_{ \mbox{c}}}
{\tan^{-1}(\mbox{e}^{-\eta^*})}.
$$

\begin{figure}[!t]
\begin{center}
\end{center}
  \caption{
Correction rate for the Sternheimer approximation with $\alpha^* \approx 0$,
 corresponding to the
 second term in the square bracket in Eq.\,(A11)
 for $\sqrt{s}$\,=\,0.1, 0.2, 0.5, 1.0\,TeV, where we assume
 $p_0 = 200$\,MeV/c for two choices of $\zeta$, 0.02 and 0.04. 
}
\end{figure}

Assuming $\bar{N}_\gamma = 2\bar{N}_{\pi^0}$, 
 and substituting the explicit form of 
$n_{\gamma}(E_{\pi^0}^*, t_{\gamma}^*$) given by
 Eq.\,(11) into Eq.\,(A10) with 
$E_{\pi^0}^* = T_{ \mbox{c}}x$,
 finally we obtain the kinematical relation between
 $\gamma$ and ${\pi^0}$ pseudo-rapidity distributions
$$
2\frac{dN_{\pi^0}}{d\eta^*} = 
\biggl [ 1 + \frac{1}{2} \biggl(\frac{m_{\pi^0}}{p_0}\biggr)^2
\frac{ {\it \Delta}(\eta^*)}{F_0(\tau_{\eta^*})}
\tanh ^2 \eta^*  \biggr]
\frac{dN_{\gamma}}{d\eta^*},
\eqno{\rm (A11)}
$$
with
$$
{\it \Delta}(\eta^*) =
\int_{x_{\eta^*}}^1
\frac{dx}{x}\biggl(\ln \frac{1+B_{x,\eta^*}}{1-B_{x,\eta^*}}
 - B_{x,\eta^*}\biggr) (1-x)^4 \mbox{e}^{-\tau_\eta^* x}
$$
$$
  \times \ 
\Biggl \{ \biggl (\frac{4/{\tau}_\eta^*}{1-x} + 1\biggr)
             \biggl (x + \frac{\zeta}{1 + \zeta \tau_\eta^*} \biggr )
             - \frac{1}{{\tau}_\eta^*}\Biggr \},
\eqno{\rm (A12)}
$$

see Eq.\ (19) for ${\tau}_\eta^*$. 
In Fig. (19), we present the numerical value of the correction rate
 for the Sternheimer approximation  with $\alpha^* \approx 0$,
 corresponding to 
the second term in the square bracket in Eq.\,(A11), where we
 assume $p_0 = 200$\,MeV/c, corresponding to approximately
 $\bar{p}_{ \mbox{T}}$\,=\,185, 189, 191, 193\,MeV/c for
 $\sqrt{s}$\,=\,0.1, 0.2, 0.5, 1.0\,TeV respectivey.  
 We find that
 it is as large as 7\% around $\eta^* \approx 2$ at $\sqrt{s} = 1$\,TeV.

\end{document}